Title

Geophysical applicability of atomic clocks: direct continental geoid mapping

Authors

Ruxandra Bondarescu [1], Mihai Bondarescu [2,3], György Hetényi [4], Lapo Boschi [5,1], Philippe Jetzer [1], Jayashree Balakrishna [6]

Affiliations

1 Institute for Theoretical Physics, University of Zürich, Zürich, Switzerland

2 University of Mississippi, Oxford, MS, USA

3 Universitatea de Vest, Timisoara, Romania

4 Swiss Seismological Service, ETH Zürich, Zürich, Switzerland

5 Institute of Geophysics, Department of Earth Sciences, ETH Zürich, Zürich, Switzerland

6 Harris-Stowe State University, St. Louis, MO, USA

Abbreviated title

Atomic clocks map the geoid directly

Corresponding author

György Hetényi, gyorgy.hetenyi@sed.ethz.ch, tel.: +41-44-632-4381, fax: +41-44-633-1065

Summary

The geoid is the true physical figure of the Earth, a particular equipotential surface of the Earth's gravity field that accounts for the effect of all sub-surface density variations. Its shape approximates best (in the sense of least squares) the mean level of oceans, but the geoid is more difficult to determine over continents. Satellite missions carry out distance measurements and derive the gravity field to provide geoid maps over the entire globe. However, they require calibration and extensive computations including integration, which is a non-unique operation. Here we propose a direct method and a new tool that directly measures geopotential differences on continents using atomic clocks. General Relativity Theory predicts constant clock rate at sea level, and faster (resp. slower) clock rate above (resp. below) sea level. The technology of atomic clocks is on the doorstep of reaching an accuracy level in clock rate (frequency ratio inaccuracy of $10^{-18}$) which is equivalent to 1 cm in determining equipotential surface (including geoid) height. We discuss the value and future applicability of such measurements including direct geoid mapping on continents, and joint gravity-geopotential surveying to invert for sub-surface density anomalies. Our synthetic calculations show that the geoid perturbation caused by a 1.5 km radius sphere with 20% density anomaly buried at 2 km depth in the Earth's crust is already detectable by atomic clocks of achievable accuracy. Therefore atomic clock geopotential surveys, used together with relative gravity data to benefit from their different depth sensitivities, can become a useful tool in mapping density anomalies within the Earth.

Word count

4268 total words in manuscript + (2 figures * 300 words) = 4868 (limited to 5000)

1. Introduction and rationale

The geopotential *U* is the potential of the Earth's gravity field. Surfaces of constant geopotential value are called equipotentials, among which the one that most closely reproduces the global mean sea surface is distinguished and is labelled geoid. This definition also provided the first approach to measure the geoid by satellite track altimetry above oceans, and later extending the geoid map over the continents (Marsh & Martin, 1982). However, fine-scale observation of the geoid is a challenging geophysical task, especially on continents. On local to regional scales, potential numbers are possible to determine from land gravity measurements and height determination (i.e., spirit levelling); however this technique also uses the derivative of the potential field. On the global scale, currently applied techniques that provide both high lateral resolution and accuracy are based on indirect approaches. Gravity measurements from several satellites became accessible in the past decade. Co-orbiting GRACE satellites (Gravity Recovery and Climate Experiment; Tapley et al., 2004; von Frese et al., 2009) provide information on the geoid through an integration of the observed gravity field. Long-term measurements of CHAMP satellite (CHAllenging Mini-satellite Payload) studied the spatial and temporal variability of the gravity and magnetic field of the Earth over an eight year period. However, their spatial resolution is about 400 km (Reigber et al. 2006). More recently, gradiometry data from GOCE satellite (Gravity Field and Steady-State Ocean Circulation Explorer) became available. Results from this mission are beginning to appear in the literature (Pavlis et al., 2008; Pail et al., 2010, 2011). To obtain the currently most accurate geoid models, two-fold integration of gradiometry measurements and calibration based on land measurements is performed (e.g., Biancale et al., 2011). In summary: to date, there is no implemented technique able to perform measurement of the geoid or any equipotential surface directly (i.e., without integration).

Here we refresh Bjerhammar's (1986) idea of relativistic geodesy at a time when a suitable tool, namely atomic clocks, are on the doorstep of reaching the necessary accuracy to measure geopotential differences directly and at a scale of interest. This tool is technically achievable in the next years, and therefore it represents a significant methodological advance that is expected to produce a new type of valuable scientific data. To anticipate and to prepare the arrival of such data, this paper's purpose is to re-introduce the theory and concept of atomic clock measurements as a way to measure geopotential differences and to discuss possible applications.

Our proposal is based on Einstein's theory of General Relativity, where massive objects curve space-time and slow down time. As a consequence, clocks beat slower near heavy objects like the Earth (Misner et al., 1973). (In more practical terms, the different tick rate of the clocks makes our legs age slower than our head by measurable amounts.) Structures below the Earth's surface with different densities with respect to their environment also affect the clock rate of local observers. A high-precision clock located over significant excess mass below the surface (e.g., large iron ore body) will beat slower than another clock located over a large cavity. In more general terms, sufficiently accurate clock frequency measurements can be directly related to differences in the geopotential of the Earth, which can then be used to infer internal density variations.

The appropriate tools for this approach are portable atomic clocks (e.g., Chou et al., 2010a; Kleppner, 2006). Experiments conducted by Chou et al. (2010a) with two optical atomic clocks connected by a telecommunication fiber wire have measured changes in clock rates due to a change in elevation of 33 cm. In the meantime, the tick rate of high-end atomic clocks is increasingly accurate (e.g., Chou et al., 2010b; Campbell et al., 2012); hence atomic clocks represent a sensitive way to directly map local changes of the geopotential with an accuracy on the order of 1 cm. Local clock measurements on continents are conceivable in a way similar to relative gravity campaigns: portable clocks will be used for surveying to deduce small scale internal structure of the Earth. The portable clock will be synchronized via a fiber wire connection to a fixed clock, which could be located at a geophysical observatory or at the mean sea level, for example. The fiber optics links have been shown to work across lengths of 250 km with inaccuracies below $10^{-18}$ in the frequency ratio (e.g., Newbury et al., 2007).

The first benefit of this approach for geophysics is a novel tool to map geopotential differences directly on continents. Compared to existing satellite-derived geoid maps, this local and direct approach will certainly add valuable details on anomalies at local and regional scales, including raw material deposits, fluid reservoirs in the crust and the internal structure of the lithosphere.

Geoid maps could be constructed from clock measurements and the directly determined geopotential differences. We conjecture that, on continents, this procedure using geopotential differences is the most direct possible form of local geoid mapping. The main advantage compared to standard methods of geoid determination from gravity measurements and/or levelling is that no integration of gravitational acceleration $g$ is necessary. Integrals of $g$, which is a vector field ($g = -\nabla U$), are poorly defined because (1) the direction of $g$ is usually not measured (in the great majority of cases only the magnitude of its vertical component is measured), and (2) more importantly, the required integration is a non-unique mathematical operation. In practical terms, equipotential surfaces can be constructed by connecting locations with the same clock rates. Then, to obtain the geoid itself, standard height reduction or downward continuation would still need to be performed, although over relatively small altitude differences.

Second, local measurements of changes in the geopotential combine well with gravitational acceleration data at the same points. These two quantities have different dependencies on the distance $R$ to a given inhomogeneity: e.g., for a spherical shape (like the Earth) the geopotential $U$ is inversely proportional to $R$, while the gravity $g$ is inversely proportional to $R^2$. The combined use of independent measurements of $U$ and $g$ anomalies provides a new way of exploring structures beneath the Earth's surface. In the case of a buried sphere of constant density anomaly, as shown by a synthetic example below, one can directly find the distance to the sphere by dividing $\Delta U$ by $\Delta g$. This example is also used to compute the minimum size of objects to which high-end atomic clocks (with an accuracy of 1 cm in equivalent geopotential height) would be sensitive as a function of depth and density anomaly. More realistic cases and surveying strategy will be presented in future work.

2. Sensitivity of atomic clocks

We here discuss future portable atomic clocks that could provide the first direct and local measurements of the geoid. For two identical clocks operating at locations $x_1$ and $x_2$, the ratio of their frequencies $f$ will depend on the geopotential $U$ of the Earth at the location of each clock. For stationary observers,

$$\frac{f(x_1)}{f(x_2)} \approx 1 - \frac{U(x_2) - U(x_1)}{c^2}, \tag{1}$$

where $c$ is the speed of light. With $\Delta U = U(x_2) - U(x_1)$, and $\Delta f = f(x_2) - f(x_1)$, the fractional frequency inaccuracy is $\Delta f / f(x_2) = \Delta U / c^2$. From Eq. 1 it can be seen that equipotential surfaces ($\Delta U = 0$) correspond to surfaces of constant clock rate ($\Delta f = 0$). If the clocks are not stationary with respect to each other, higher order corrections need to be included (Blanchet et al., 2001).

Some of the first calculations for the relativistic rate shift of clocks in the vicinity of the Earth, including all terms larger than one part in $10^{18}$, were performed by Wolf & Petit (1995), who also discussed the synchronization of clocks (Petit & Wolf, 1994). However, the frequency ratio inaccuracy of the best atomic clocks of the time was only about $10^{-15}$. The most accurate atomic clocks to date have reached a frequency inaccuracy of $8.6 \times 10^{-18}$ (Chou et al., 2010b). Technology that could improve the frequency inaccuracy of atomic clocks to $10^{-19}$ is currently being tested (Campbell et al., 2012). These highly accurate clocks are still laboratory size devices. A compact, but less accurate clock with $\Delta f / f \sim 10^{-16}$ is already built for the Atomic Clock Ensemble in Space mission (Heβ et al., 2011). Progress is being made towards increasing the accuracy to $\Delta f / f \sim 10^{-18}$, which is the envisioned sensitivity for clocks that are being built for the Space-Time Explorer and Quantum Test of the Equivalence (STE-QUEST), one of the four possible medium size missions to be implemented in European Space Agency (ESA)'s Cosmic Vision program in 2022. The synchronization of portable atomic clocks through space is a possibility, either between a clock on land and a clock on a satellite, or between two clocks on land through a telecommunication satellite. However, the accuracy of such communications can be a problem as they may be affected by atmospheric turbulences (e.g., Djerroud et al., 2010).

In the following calculations we assume a frequency inaccuracy $\Delta f / f$ of $\sim 10^{-18}$. This frequency inaccuracy corresponds to a change in the gravitational potential of $\Delta U = c^2\, 10^{-18} \approx 0.1$ m² s⁻². The corresponding sensitivity in geoid height can be computed by assuming the two clocks are at height $R_E + \Delta h$ and $R_E$, respectively, leading to

$$\Delta U = \frac{GM_E}{R_E} - \frac{GM_E}{R_E + \Delta h} \approx \frac{GM_E \Delta h}{R_E^2}, \qquad (2)$$

where $G$ is Newton's gravitational constant, $R_E$ and $M_E$ are the mean radius and mass of the Earth. For the standard values of $R_E \approx 6371$ km and $M_E \approx 5.97 \times 10^{24}$ kg, and $\Delta U = 0.1$ m² s⁻², one obtains a sensitivity in geoid height of $\Delta h \approx 1$ cm.

3. Applicability to direct geoid measurements

The geoid over oceans can be known at relatively high (between minute and degree) spatial resolution. While the oceanic crust is thought to be simpler both in structure and in density variations than the continental crust, the amplitude of geoid anomalies in the ocean due to the underlying mantle structure reach up to a height of 100 m (Marsh & Martin, 1982). On continents, a resolution of 1 cm in geoid height with unlimited spatial resolution, which is achievable considering current technological advances, will therefore be largely sufficient to add significant details to existing geoid maps derived from satellite measurements. These direct measurements of geopotential differences, leading to a more direct continental geoid as outlined above (section 1), will not only provide the first, high-resolution geoid map *per se*, but will also reveal density anomalies from large to small, mantle to crustal scale sources, including structure of the lithosphere, fluid reservoirs in the crust and dense ore deposits. Moreover, inverse method approaches looking to map structural and/or density variations of the Earth's interior will further benefit from the joint geopotential and gravity surveys.

4. Geopotential and gravity anomalies

While gravimeters (both absolute and relative) have become highly accurate instruments, gravity alone is insufficient to map sub-surface structures. This is due to the non-uniqueness of the inverse problem. Combining local gravity with direct, local geopotential measurements reduces some of this degeneracy (i.e., depth determination) and is able to provide meaningful inversion results. In practical terms, using already available relative gravimeters and future portable atomic clocks, joint field surveying becomes a conceivable approach.

In this study, we take two simple synthetic examples for a joint gravity-geopotential survey: a buried sphere (see below) and a buried 3D rectangular slab (see Appendix A), both of constant density anomaly. We then investigate the extent to which the degeneracy of the inverse problem is reduced by having both gravity and geopotential measurements.

4.1 Buried sphere

We assume a spherical body of density $\rho_1$ buried at a depth $h$ in a medium of density $\rho_0$. The corresponding geopotential anomaly (Turcotte & Schubert, 2002) is

$$\Delta U(x) = \frac{4\pi G b^3 \Delta \rho}{3(x^2 + h^2)^{1/2}}, \tag{3}$$

where $\Delta\rho = \rho_1 - \rho_0$ is the density anomaly of the sphere compared to its surroundings, $b$ is its radius, and $x$ is horizontal distance on the surface measured from the centre of the sphere. The related gravity anomaly in the $z$-direction is

$$\Delta g_z = \frac{4\pi G h b^3 \Delta \rho}{3(x^2 + h^2)^{3/2}}. \tag{4}$$

As a synthetic example, we consider a spherical (radius $b$ = 1.5 km) piece of mantle or magma ($\rho_1 \approx$ 3200 kg m$^{-3}$) located in the shallow upper crust ($\rho_1 = \rho_{crust} \approx$ 2670 kg m$^{-3}$) at a depth of $h$ = 2 km, which is a relative density anomaly of $\Delta\rho$ = 20% $\rho_{crust}$. Such a scenario (Fig.1) would easily be detectable with both atomic clocks ($\Delta U_{max} \approx$ 0.25 m$^2$ s$^{-2}$) and relative gravimeters ($\Delta g_{z\ max} \approx$ 12.6 mGal). The primary limitation comes from the atomic clock sensitivity since relative gravimeters have sensitivities in the range of µGal. We then consider the ratio of $\Delta U$ and $\Delta g_z$ as a function of $x$:

$$\frac{\Delta U}{\Delta g_z} = h + \frac{x^2}{h} \approx h \text{ for } x << h. \tag{5}$$

In this simple case, the above ratio can be used to determine the depth $h$ to the centre of the anomaly (see Fig.1b).

To illustrate the sensitivity of our method, we compute from Eq. 3 the radius of the smallest detectable sphere as a function of depth, density contrast and clock sensitivity:

$$b = \left(\frac{3h \Delta U_{max}}{4\pi G \Delta \rho}\right)^{1/3}, \tag{6}$$

where $\Delta U_{max} = \Delta U(x=0)$.

Figure 2 displays $b$ for density contrast ratios $\Delta\rho / \rho_{crust}$ ranging from 1% to 20% as a function of depth. The amplitude of eventual errors or noise is not yet known, but for safety we use a frequency inaccuracy ratio two times higher than expected ($\Delta U_{max} \approx$ 0.2 m$^2$ s$^{-2}$). Figure 2 shows the dashed lines above which atomic clocks are able to detect buried objects for each density contrast. Only buried ($h \geq b$) spheres are considered. At large $\Delta\rho / \rho_{crust}$ of 20%, a ~ 4 km radius sphere is detectable down to 45 km depth. For a lower density contrast of 1%, a 10 km sphere can be measured to a depth of ~ 37 km. For a 3D example, we refer to Appendix A.

For more complex structures or future real surveys, equations 5 and 6 cannot be expressed analytically, but the different sensitivities of gravity (proportional to $1/R^2$) and geopotential (proportional to $1/R$) measurements will still reduce the degeneracy of the inverse problem compared to the case when only gravity is used. It is also important to note that equations 3 and 4 have different physical meanings: due to the relative nature of easily performed gravity measurements in the field, as well as to the inherent properties of integration, the two methods will provide independent measurements and it is their different sensitivity that will allow better determination of sub-surface structures.

5. Conclusions

Mapping the geoid directly and locally with portable atomic clocks is a dramatic application of General Relativity to our everyday life, which is becoming possible with improving technology. State of the art atomic clocks are expected to reach sensitivities of the order of $10^{-18}$ in frequency ratio inaccuracy within the next decade; this corresponds to a sensitivity of $\Delta h = 1$ cm in geoid height and to a geopotential difference of $\Delta U = 0.1$ m$^2$ s$^{-2}$. It will be soon possible to conduct measurements with a portable atomic clock around the Earth in the same way as with a relative gravimeter, and the spatial resolution of geopotential measurements would become unlimited. Such atomic clocks could be used to add significant details to current geoid maps derived from satellite measurement and extensive computations.

In the temporary lack of real data from high accuracy atomic clock measurements, we computed synthetic examples to demonstrate the interest and applicability of the approach. Using the simple case of a buried sphere of constant density anomaly, we explored the minimum detection size of spheres to which atomic clocks of achievable accuracy would be sensitive. The radius of the object to detect $b$ is proportional to $h^{1/3}\Delta\rho^{-1/3}$; that is, for example, at 2 km depth and 20% density anomaly, a sphere of 1.5 km radius could be detected; or, a larger object of 4 km radius and the same density contrast is possible to detect down to 45 km depth. At the moment we cannot yet quantify the magnitude of practical errors (as the current accuracy of portable tools is 100 times lager), but the above calculations were made by considering a two-fold safety margin of the expected accuracy. Thus, the potential for a powerful tool is present, and the approach is to be validated upon completion of the instruments.

We also proposed to combine local gravity measurements with local geopotential measurements from atomic clocks. Since the gravitational acceleration and the potential have different dependencies of the distance to the anomaly, combining these measurements adds information towards imaging density structure at depth. In the case of the buried sphere of constant density, one can find the distance $h$ to the buried structure from $\Delta U / \Delta g_z$, which is parabola with a value of $h$ at its apex. In more realistic examples the shape and the amplitude of the density heterogeneity are unknown. However, having measurements of both $\Delta U$ and $\Delta g_z$ reduces the degeneracy of the inverse problem in mapping underground structures. Practical applications may require area-wise mapping of both geopotential and gravity.

Atomic clocks are becoming compact, portable and stable enough to fly in space. ESA has already built a compact atomic clock with an accuracy of $10^{-16}$ that will be placed on the International Space Station by 2014 (Heβ et al., 2011). Clocks with accuracies more than an order of magnitude higher exist in laboratories, and even more accurate clocks are being developed. We believe that this is the right time to start a discussion in the geophysics community about possible benefits from this new technology. Other potential applications include using successive measurements of geopotential and gravity to understand underground water or magma movements. Such measurements might also be used in post-earthquake analysis to map the related structural changes. Local and direct measurements of the geoid will likely have significant impact on the current mapping of Earth structure on continents.

Acknowledgments

We thank David Hume, Steve Lecomte, Andrew Lundgren, Domenico Giardini and Alain Geiger for useful discussions. We also thank two anonymous reviewers and the Editor for their comments and advices. RB and PJ acknowledge support from the Dr. Tomalla Foundation and the Swiss National Foundation.

Figure legends

Figure 1. (a) Geopotential anomaly $\Delta U$ and gravity anomaly $\Delta g$ as a function of displacement for a sphere of radius $b = 1.5$ km and a density contrast of 20% with respect to the surrounding crust buried with its centre at a depth of $h = 2$ km. The red line shows the detection threshold of $\Delta U_{max} = 0.1$ m$^2$ s$^{-2}$ for atomic clocks with frequency inaccuracies of $10^{-18}$. (b) To determine the depth to the centre of the sphere, $\Delta U / \Delta g$ is displayed as function of horizontal position. The resulting curve is a parabola with its minimum located at the searched depth $h$.

Figure 2. Minimum radius of a buried sphere of radius $b$ that atomic clocks are able resolve, as a function of depth $h$ and relative density anomaly $\Delta \rho / \rho$ of 1, 2, 5 and 20%. Here we use an accuracy of $\Delta U_{max} = 0.2$ m$^2$ s$^{-2}$, that is the double of the expected accuracy, in order to take into account potential measurement errors. The upper left part of the diagram, shaded in brown, corresponds to spheres touching the ground; therefore the detection threshold curves are not calculated for this domain.

APPENDIX A

Similarly to the synthetic example of a buried sphere in section 4.1, here we present the case of a buried 3D rectangular body (slab) of constant density. This could be a simple approximation for a fluid such as a water nappe or a hydrocarbon reservoir. The relative gravity and geopotential anomalies due to a buried parallelepiped with corners ($x_i$, $y_j$, $z_k$) can be derived analytically (Banerjee & Das Gupta, 1977; Coggon, 1976; Parasnis, 1977) to be

$$\Delta g_z = G \int_{x_1}^{x_2} dx \int_{y_1}^{y_2} dy \int_{z_1}^{z_2} dz \frac{z}{(x^2+y^2+z^2)^{3/2}} \Delta\rho$$

$$= G\Delta\rho \sum_{i,j,k=1}^{2} (-1)^{i+j+k} \left[ z_k \arctan \frac{x_i y_j}{z_k \sqrt{x_i^2+y_j^2+z_k^2}} - x_i \log(y_j + \sqrt{x_i^2+y_j^2+z_k^2}) \right.$$
$$\left. - y_j \log(x_i + \sqrt{x_i^2+y_j^2+z_k^2}) \right]$$
(A1)

and

$$\Delta U = G \int_{x_1}^{x_2} dx \int_{y_1}^{y_2} dy \int_{z_1}^{z_2} dz \frac{1}{\sqrt{x^2+y^2+z^2}} \Delta\rho$$

$$= G\Delta\rho \sum_{i,j,k=1}^{2} (-1)^{i+j+k} \left[ -\frac{z_k^2}{2} \arctan \frac{x_i y_j}{z_k \sqrt{x_i^2+y_j^2+z_k^2}} - \frac{y_j^2}{2} \arctan \frac{x_i z_k}{y_j \sqrt{x_i^2+y_j^2+z_k^2}} \right.$$
$$-\frac{x_i^2}{2} \arctan \frac{y_j z_k}{x_i \sqrt{x_i^2+y_j^2+z_k^2}} + x_i y_j \log(z_k + \sqrt{x_i^2+y_j^2+z_k^2}) + y_j z_k \log(x_i + \sqrt{x_i^2+y_j^2+z_k^2})$$
$$\left. + x_i z_k \log(y_j + \sqrt{x_i^2+y_j^2+z_k^2}) \right],$$
(A2)

where $\Delta\rho$ is the density contrast with the surrounding material.

We find that $\Delta U / \Delta g_z$ is a linear function of $z_1$ for a body of fixed thickness $z_2 - z_1$. Figure 3 shows $\Delta U / \Delta g_z$ above the centre of the body, where both $\Delta U$ and $\Delta g_z$ are maximum, using a relative density anomaly if $\Delta\rho / \rho_{crust}$ = 2%. If the horizontal extent of the body ($x_1$, $x_2$, $y_1$ and $y_2$) are determined from e.g., gravity gradient measurements, then one can constrain $z_2$ and $z_1$ from $\Delta U$ and $\Delta g_z$ measurements. The slope of the line approaches zero as the size of the slab becomes very large in the x- and y-directions. This is consistent with the infinite slab approximation where both $\Delta U$ and $\Delta g_z$ are proportional to the thickness of the slab, but independent of depth.

To test the sensitivity to this type of object we take a maximum of $\Delta U_{max} = 0.2$ m$^2$ s$^{-2}$ to account for eventual measurement errors, $\Delta\rho = 0.02\ \rho_{crust}$ with $\rho_{crust} = 2670$ kg m$^{-3}$ and a thickness of $z_2 - z_1 = 2.5$ km. This way we find that atomic clocks could be sensitive to a 10 km by 10 km slab at a depth to the top of $z_1 = 1.5$ km or less, or to a 10 km by 40 km body down to a depth of $z_1 \sim 13$ km.

Figure legends

Figure 3. The value of $\Delta U / \Delta g_z$ above the centre of the 3D rectangular body as a function of the depth to the body's top $z_1$ for a body thickness of $z_2 - z_1 = 2.5$ km. In the two cases, the horizontal size of the body is $\Delta x = x_2 - x_1 = 10$ km and $\Delta y = y_2 - y_1 = 40$ km (solid red line), and then $\Delta x = \Delta y = 40$ km (dashed blue line). While in the second (blue) case $\Delta U / \Delta g_z$ varies only about 150 m with varying $z_1$, it varies about 1 km in the first (red) case for the same change in depth. This is consistent with the expected weakening of depth-dependency with horizontally increasing size of the slab.

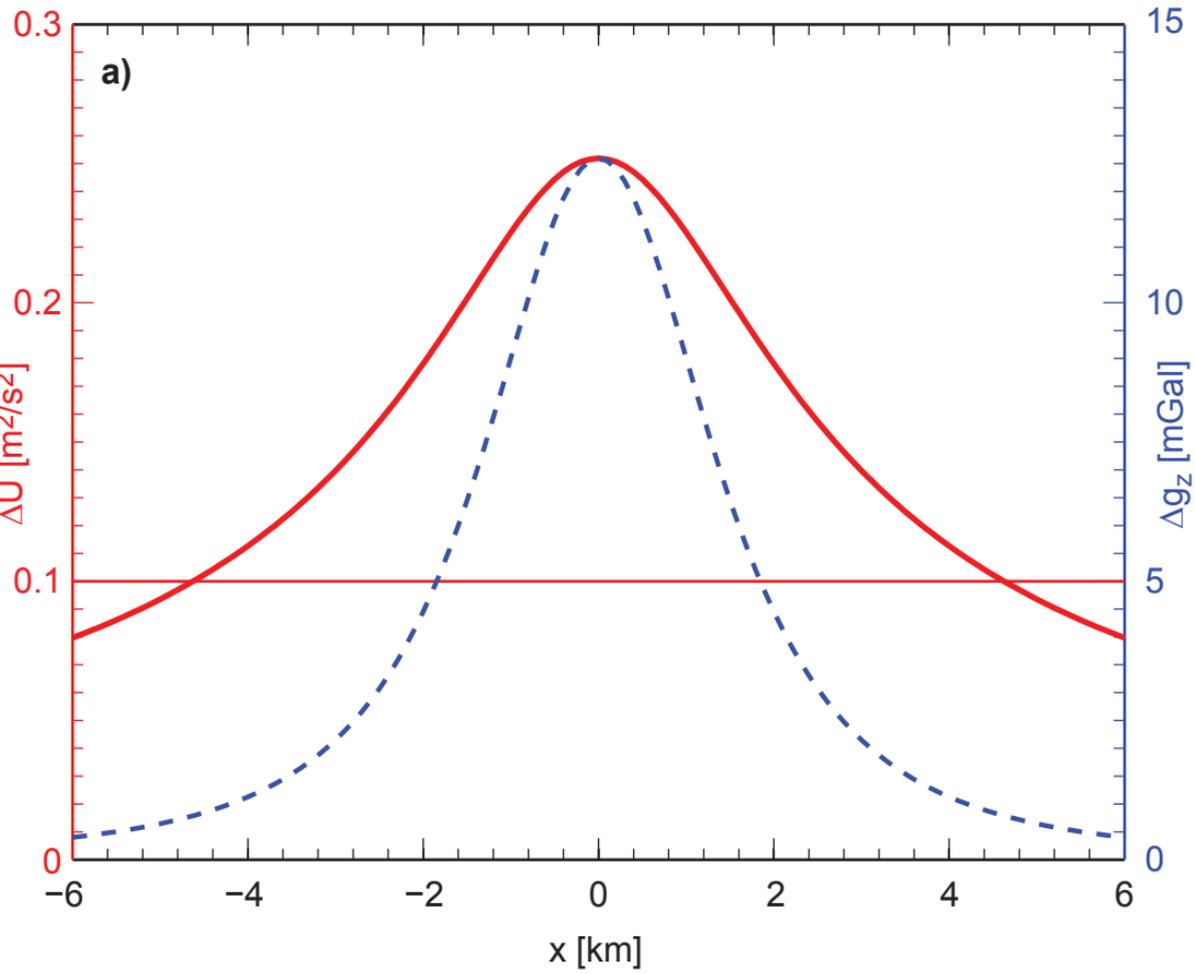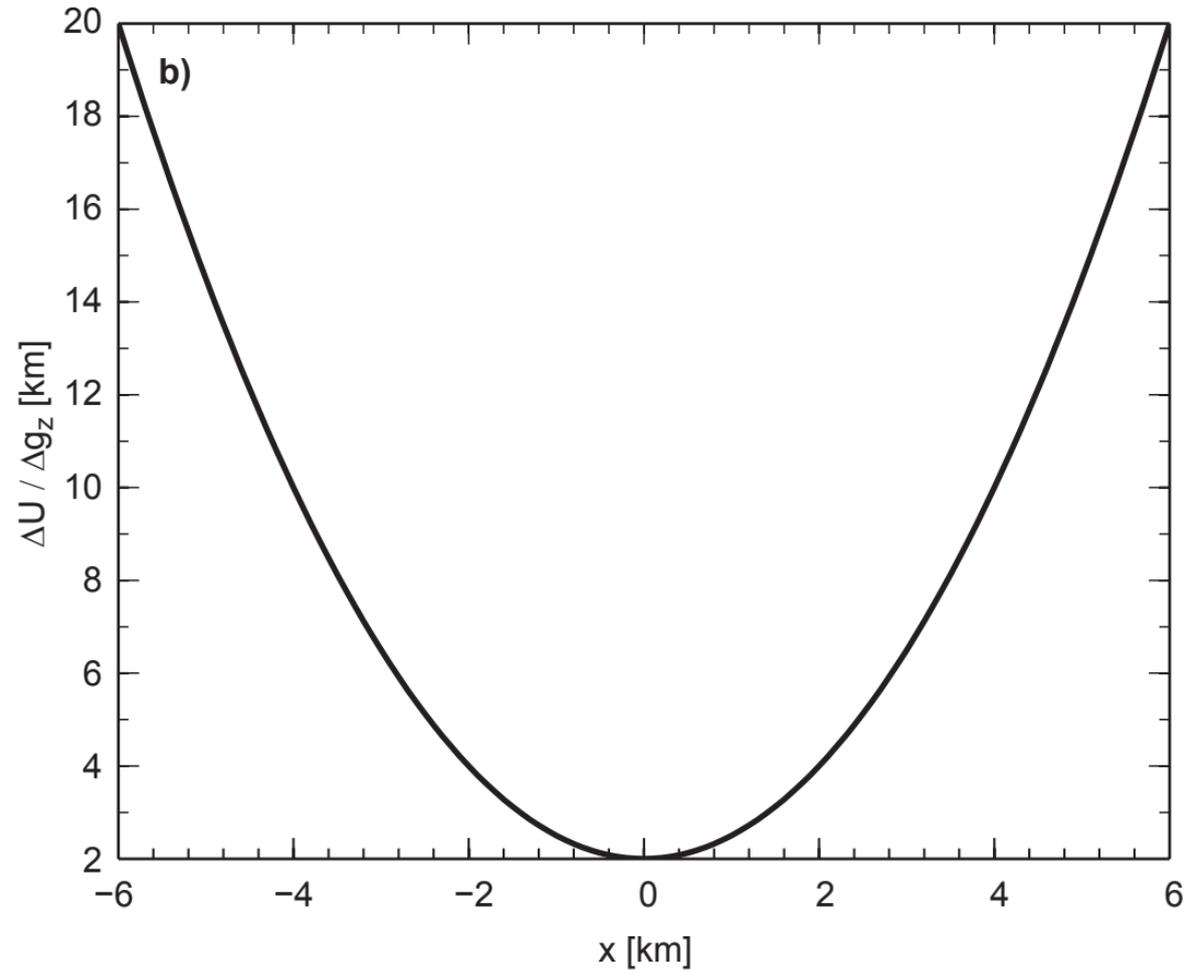

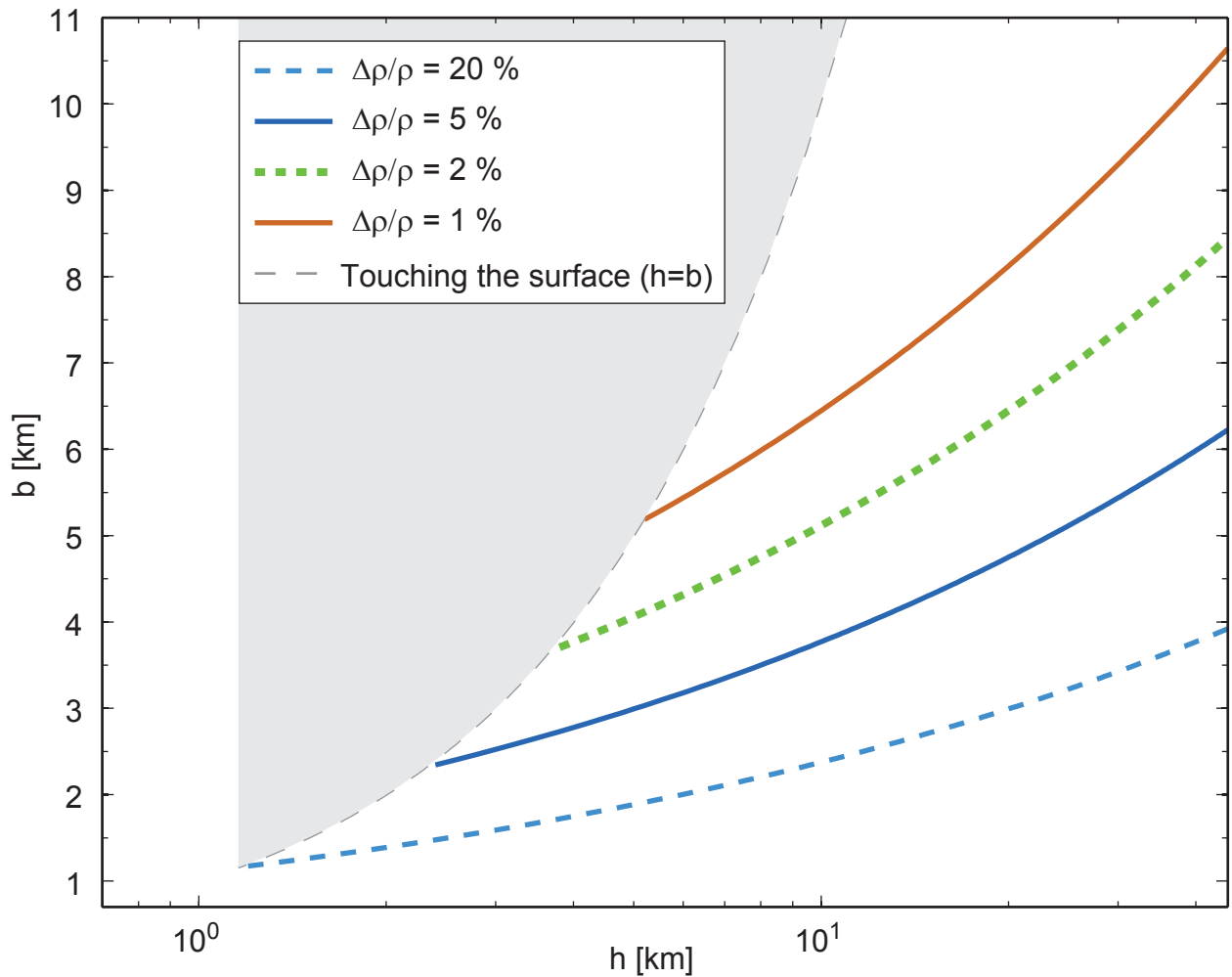

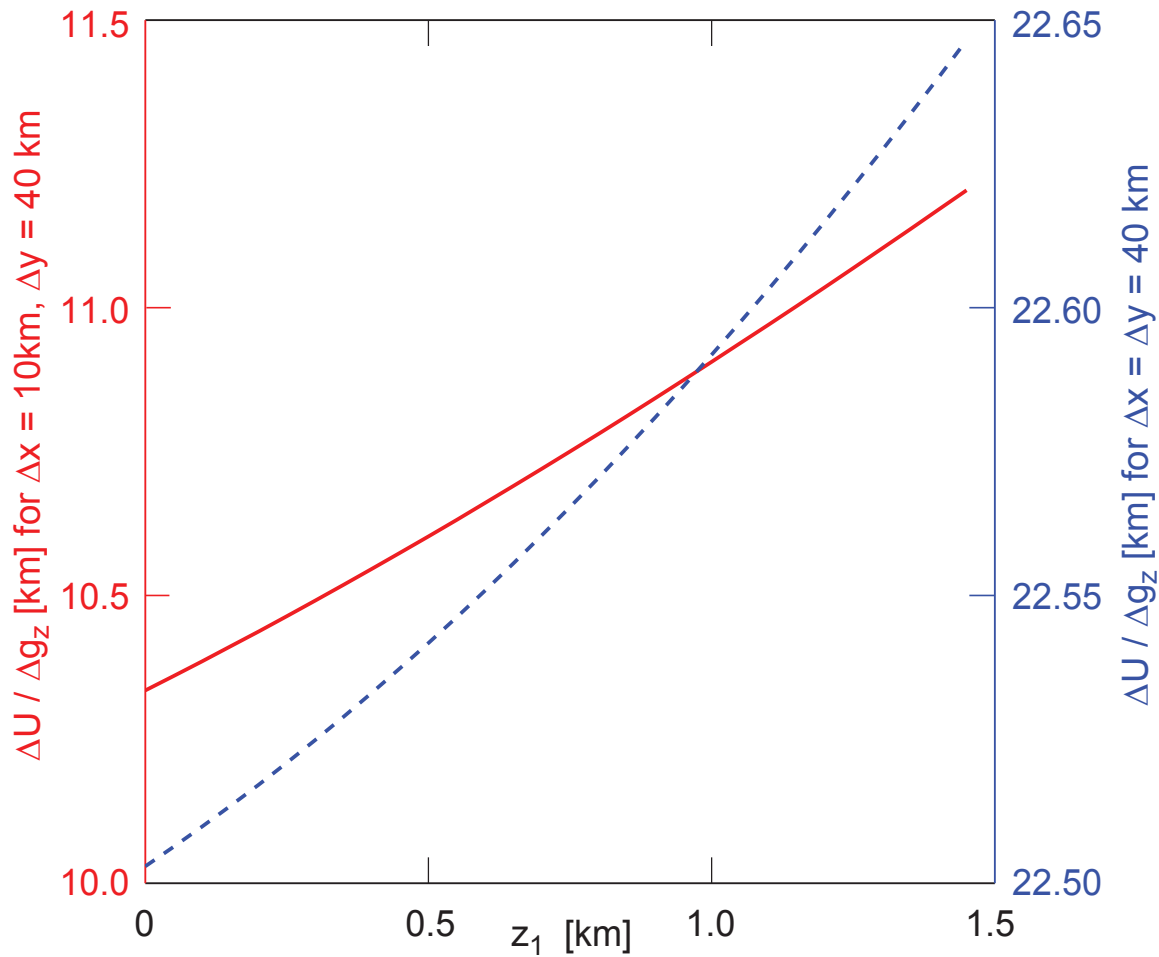